\documentclass[twocolumn]{aastex702}

\graphicspath{{./}}
\usepackage{amsmath}
\usepackage{placeins}

\received{August 13, 2026}
\accepted{September 15, 2026}

\begin{document}

\title{A Formation-stage Bottleneck for Exomoons around Close-in Rocky Planets}

\correspondingauthor{Hongping Deng}

\author[0009-0003-1435-4977]{Rongxi Bi}
\affiliation{Research Center for Planetary Science, College of Earth and Planetary Sciences, Chengdu University of Technology, No. 1, East Third Road, Erxianqiao, Chenghua District, Chengdu 610059, China}
\email{rongxbi@stu.cdut.edu.cn}

\author[0000-0001-6858-1006]{Hongping Deng}
\affiliation{Shanghai Astronomical Observatory, Chinese Academy of Sciences, 80 Nandan Road, Shanghai 200030, China}
\email[show]{hpdeng353@shao.ac.cn}

\author[0000-0002-4535-3956]{Christian Reinhardt}
\affiliation{Department of Astrophysics, University of Zurich, Winterthurerstrasse 190, CH-8057 Zurich, Switzerland}
\affiliation{Physics Institute, Space Research and Planetary Sciences, University of Bern, Sidlerstrasse 5, CH-3012 Bern, Switzerland}
\email{christian.reinhardt@uzh.ch}

\author[0000-0002-9442-137X]{Shangfei Liu}
\affiliation{School of Physics and Astronomy, Sun Yat-sen University, No. 2 Daxue Road, Tangjiawan, Xiangzhou District, Zhuhai, Guangdong 519082, China}
\email{liushangfei@mail.sysu.edu.cn}

\author[0000-0003-1000-143X]{Yun Liu}
\affiliation{Research Center for Planetary Science, College of Earth and Planetary Sciences, Chengdu University of Technology, No. 1, East Third Road, Erxianqiao, Chenghua District, Chengdu 610059, China}
\affiliation{State Key Laboratory of Critical Mineral Research and Exploration, Institute of Geochemistry, Chinese Academy of Sciences, 99 West Lincheng Road, Guanshanhu District, Guiyang, Guizhou 550081, China}
\email{Liuyun@vip.gyig.ac.cn}

\begin{abstract}
Massive moons around rocky exoplanets are expected outcomes of giant impacts, yet no exomoon has been securely confirmed. We address this problem using meshless finite-mass (MFM) simulations of giant impacts between differentiated rocky planets that include stellar tidal and Coriolis forces in a local corotating frame for orbital periods of 1--300 days, and compare these simulations with otherwise identical collisions in isolation. We find that stellar perturbations impose a severe formation-stage bottleneck on moon-forming disks. Ultra-short-period impacts leave essentially no surviving circumplanetary disk (CPD), whereas 10-day systems retain only strongly depleted, radially truncated, and dynamically compact disks. High impact velocities, expected for close-in planets, are devastating for disk generation, while retrograde impacts can slightly compensate for disk mass depletion. These results show that stellar perturbations can suppress exomoon formation around close-in rocky planets, reshaping expectations for the demographics of exomoons.
\end{abstract}


\section{Introduction}

Natural satellites provide key constraints on planet formation \citep{Knierim2026Jupiter,fang2025moon}, planetary spin--orbit evolution \citep{touma1994evolution,farhat2022resonant,su2026coupled}, and planetary habitability. In the solar system, the Moon has been argued to stabilize Earth's obliquity, thereby reducing chaotic variations in insolation and climate \citep{laskar1993stabilization,Li2014ApJ}. Yet, despite the discovery of thousands of exoplanets by transit surveys such as Kepler and TESS, no exomoon has been securely confirmed \citep{kipping2022exomoon,heller2024large,millholland2025exploring}.

This observational deficit contrasts with expectations from the solar system, where satellites are ubiquitous and arise through diverse formation pathways \citep{peale1999origin,jewitt2007irregular}. Regular satellites of gas giants primarily form in gaseous circumplanetary disks (CPDs) \citep{canup2006common,canup2009origin}, whereas massive moons around rocky and icy bodies are typically linked to giant impacts. Here we focus on exomoons around rocky planets, which are potentially habitable and are primary targets of next-generation transit surveys such as PLATO \citep{rauer2018space,rauer2025plato} and Earth 2.0 \citep{ge2022mission}. For these worlds, giant impacts are a leading mechanism for generating satellite-forming debris disks and thus moons, as exemplified by the Earth--Moon \citep{canup2004simulations,cuk2012making,deng2019primordial} and Pluto--Charon systems \citep{mckinnon1989origin,canup2005giant,canup2011giant,Denton2025}. We note that massive exomoons, or even binary planets, may also form via tidal capture during planetary encounters \citep{deng2020hypothetical}.

Giant impacts are expected to be common during late-stage terrestrial planet accretion, when planetary embryos undergo orbit crossing, scattering, and stochastic collisions \citep{morbidelli2012building,chambers2013late}. This mechanism is also relevant to close-in super-Earths, which constitute a major fraction of the known close-in exoplanet population \citep{howard2012planet,petigura2013prevalence,hsu2019occurrence}. One pathway to late giant impacts in these systems is the breaking of compact resonant chains assembled through convergent migration. After gas-disk dispersal, many such chains become dynamically unstable, driving orbit crossing and giant impacts \citep{izidoro2017breaking}. The relative scarcity of resonant architectures among mature systems is consistent with this evolutionary pathway \citep{dai2024prevalence}.

If giant impacts are expected, the absence of confirmed exomoons around super- and sub-Earths is likely due to three factors: (1) challenges in observing exomoons, (2) difficulty in retaining exomoons, and (3) inefficient moon formation around close-in rocky planets. The observational bias is difficult to quantify given the lack of even a single confirmed exomoon. Various exomoon detection methods have been proposed, including transit radius variations (TRV) and transit timing variations (TTV), but detecting exomoons around rocky planets remains challenging even for next-generation transit surveys \citep[e.g.,][and references therein]{rodenbeck2020exomoon,cui2026simulation}. Nevertheless, a recent study suggests that JWST may detect exomoons larger than Ganymede around terrestrial planets \citep{pass2026jwst}.

On the theoretical side, the survival of moons has been extensively studied \citep{barnes2002stability,sasaki2012outcomes,dobos2021survival,patel2026tidally}. In general, stable satellite orbits occupy only a fraction of the planetary Hill sphere, with the stable region shrinking for eccentric planetary or satellite orbits \citep{domingos2006stable}. In addition, the spin of the host planet appears to be a key factor in determining an exomoon's survival. When the planetary spin rate is lower than the moon's orbital mean motion, planetary tides drive inward migration, as observed for Phobos \citep{Bagheri2021}, which may eventually reach the Roche limit and undergo tidal disruption. In contrast, a fast-spinning planet can drive outward tidal migration, eventually causing the moon to escape the planet's stable region \citep{barnes2002stability,sasaki2012outcomes}.
It has been shown that moon survivability increases for long-period planets, and that close-in planets (orbital periods $< 10$ days) generally retain no long-lived moons \citep{dobos2021survival}. However, such constraints apply only after a moon has already formed. They do not address the prerequisite formation-stage question considered here: whether debris generated by a giant impact around a close-in rocky planet remains bound and retains sufficient angular momentum to form a satellite-forming disk.

Here we focus on the formation process itself and, more specifically, on the masses of satellite-forming disks expected from giant impacts in the current transiting rocky-planet population. Previous numerical simulations across different planetary scales have shown that such collisions can, in principle, generate disks massive enough to accrete detectable moons \citep{barr2017formation,nakajima2022large}.
For the many known rocky exoplanets on close-in orbits, however, the local collision environment differs qualitatively from that of the solar system: impact debris evolves in a noninertial, corotating environment shaped by stellar tides and Coriolis forces \citep{murray1999solar}. Previous hydrodynamic simulations have shown that stellar tides can influence the postimpact evolution of close-in, volatile-rich super-Earths by promoting the loss of extended H/He envelopes beyond the Hill sphere \citep{liu2015giant}. However, whether and to what extent the same close-in tidal environment suppresses the production of satellite-forming disks remains unclear.

In this work, we test whether giant impacts onto close-in rocky planets can generate and retain satellite-forming debris disks in the presence of stellar perturbations. We introduce a giant-impact simulation framework that self-consistently includes the stellar tidal field and Coriolis forces (Section~\ref{sec:method}) and compare each close-in impact with an otherwise identical impact evolved in isolation. Using the surviving bound mass and its angular-momentum distribution, we quantify how the close-in orbital environment regulates disk formation as a function of orbital period (Section~\ref{sec:results}). We discuss the disk-retention scaling for M-dwarf systems and the role of resonant-chain breaking in exomoon formation (Section~\ref{sec:discussion}). These results place a formation-stage constraint on the expected occurrence and detectability of moons around close-in terrestrial exoplanets (Section~\ref{sec:conclusion}).

\section{Method}
\label{sec:method}

\subsection{Numerical Model and Stellar Tidal Field}

We performed giant-impact simulations with the meshless finite-mass (MFM) method implemented in the multimethod GIZMO code \citep{hopkins2015new}. MFM is a hybrid Godunov-type Lagrangian method that updates fluid variables using fluxes exchanged between effective volumes, similar to finite-volume methods. As a result, it does not rely on artificial viscosity for shock capturing and is less prone to unphysical dissipation than traditional smoothed particle hydrodynamics (SPH) methods \citep[see, e.g.,][and references therein]{cabezon2026particles}. In giant-impact simulations, MFM better captures mixing induced by subsonic turbulence than the SPH implementation in GIZMO, while using the same gravity solver and material properties \citep{deng2019enhanced}. We modeled the thermodynamic response of iron and silicate materials during shock compression and subsequent release using the ANEOS equation of state (EOS) \citep{Thompson_EOS,MELOSH_EOS,stewart2020shock}, with tabulated EOSs for an iron core and a forsterite mantle \citep{stewart2019forsterite,stewart2020iron}. GIZMO, in both MFM and SPH modes, has been well tested and widely applied to giant-impact and Moon-formation simulations \citep{deng2019primordial,deng2019enhanced,yuan2023moon,zhou2024research,fang2025moon}.

To incorporate perturbations from the host star, we adopted a local corotating frame centered on the protoplanet, similar to the shearing-box approximation but without shearing periodic boundary conditions along the radial coordinate \citep{goldreich1965ii,deng2019local}. The impact occurs on a scale much smaller than the heliocentric distance $a$, i.e., $|x|, |z| \ll a$, where $x$ and $z$ denote the local radial and vertical coordinates, respectively. Expanding the stellar gravitational potential around the planetary orbital radius $a$ and including the centrifugal acceleration in the corotating frame yields the linearized effective tidal acceleration:
\begin{equation}
-\nabla \Phi_{\text{tidal}}
= 2q\Omega^2 x \, \hat{\mathbf{x}} - \Omega^2 z \, \hat{\mathbf{z}},
\end{equation}
where $\Omega\equiv|\Omega_z|$ is the planet's orbital angular frequency and $\boldsymbol{\Omega}=\Omega_z\hat{\mathbf{z}}$ is the angular-velocity vector of the corotating frame. We define the positive $z$-direction to be aligned with the angular momentum of the prescribed impact trajectory, which also corresponds to the angular-momentum direction of the resulting debris disk. For a Keplerian orbit, the shear parameter is $q=3/2$.

The equation of motion therefore includes the hydrodynamic pressure-gradient force, self-gravity, the Coriolis force, and the effective stellar tidal force:
\begin{equation}
\label{eq:equation_of_motion}
\frac{D\mathbf{u}}{Dt}
= -\frac{1}{\rho}\nabla P
+ \mathbf{a}_{\text{sg}}
- 2\boldsymbol{\Omega} \times \mathbf{u}
+ 2q\Omega^2 x \, \hat{\mathbf{x}}
- \Omega^2 z \, \hat{\mathbf{z}},
\end{equation}
where $\mathbf{a}_{\rm sg}$ is the gravitational acceleration generated by the planetary bodies and debris, calculated with a tree algorithm \citep{hopkins2015new}.

We also include the work done by the tidal field in the total energy equation:
\begin{equation}
\frac{\partial \rho e}{\partial t}
+ \nabla\cdot\left[(\rho e+P)\mathbf{u}\right]
= \rho\, \mathbf{u}\cdot\mathbf{a}_{\text{sg}} - \rho\,\mathbf{u}\cdot\nabla\Phi_{\text{tidal}}.
\end{equation}
Stellar perturbations are strongest for impacts onto close-in planets, where $\Omega$ is large. Including these stellar forces is the main difference between this work and previous giant-impact simulations of exomoon formation \citep[e.g.,][]{barr2017formation,nakajima2022large}.

\subsection{Initial Conditions and Impact Setup}
\label{sec:impact_setup}
 
\begin{deluxetable*}{lcccccccccccc}
\tabletypesize{\scriptsize}
\tablewidth{0pt}
\tablecaption{Summary of fiducial impact suites. \label{tab:fiducial_suites}}
\tablehead{
\colhead{ID} &
\colhead{$M_{\rm tar}$} &
\colhead{$R_{\rm tar}$} &
\colhead{$M_{\rm imp}$} &
\colhead{$R_{\rm imp}$} &
\colhead{$v_{\rm imp}$} &
\colhead{$\beta$} &
\colhead{$P_{\rm orb}$} &
\colhead{$M_{\rm tot}$} &
\colhead{$M_{\rm disk}$} &
\colhead{$M_{\rm planet}$} &
\colhead{$M_{\rm esc}$} &
\colhead{$R_{\rm Hill}/R_{\rm planet}$}
}
\startdata
\multicolumn{13}{c}{Graze-and-merge impact suite: $v_{\rm imp}=1.0v_{\rm esc}$} \\
\hline
G-1 & 0.90 & 1.0 & 0.21 & 0.65 & 1.0 & 45 & $\infty$ & 1.11 & 1.42 & 1.07 & 0.02 & $\infty$ \\
G-2 & 0.90 & 1.0 & 0.21 & 0.65 & 1.0 & 45 & 300 & 1.11 & 1.52 & 1.07 & 0.02 & 207.09 \\
G-3 & 0.90 & 1.0 & 0.21 & 0.65 & 1.0 & 45 & 100 & 1.11 & 0.84 & 1.08 & 0.02 & 99.56 \\
G-4 & 0.90 & 1.0 & 0.21 & 0.65 & 1.0 & 45 & 30 & 1.11 & 1.09 & 1.07 & 0.03 & 44.62 \\
G-5 & 0.90 & 1.0 & 0.21 & 0.65 & 1.0 & 45 & 10 & 1.11 & 0.40 & 1.07 & 0.03 & 21.45 \\
G-6 & 0.90 & 1.0 & 0.21 & 0.65 & 1.0 & 45 & 1 & 1.11 & 0.00 & 0.93 & 0.18 & 4.62 \\
G-7 & 0.90 & 1.0 & 0.21 & 0.65 & 1.0 & 60 & $\infty$ & 1.11 & 3.16 & 1.05 & 0.02 & $\infty$ \\
G-8 & 0.90 & 1.0 & 0.21 & 0.65 & 1.0 & 60 & 300 & 1.11 & 3.28 & 1.05 & 0.02 & 207.09 \\
G-9 & 0.90 & 1.0 & 0.21 & 0.65 & 1.0 & 60 & 100 & 1.11 & 2.16 & 1.05 & 0.04 & 99.56 \\
G-10 & 0.90 & 1.0 & 0.21 & 0.65 & 1.0 & 60 & 30 & 1.11 & 2.64 & 1.05 & 0.03 & 44.62 \\
G-11 & 0.90 & 1.0 & 0.21 & 0.65 & 1.0 & 60 & 10 & 1.11 & 0.43 & 1.08 & 0.02 & 21.45 \\
G-12 & 0.90 & 1.0 & 0.21 & 0.65 & 1.0 & 60 & 1 & 1.11 & 0.00 & 0.91 & 0.20 & 4.62 \\
G-13 & 3.0 & 1.4 & 0.63 & 0.89 & 1.0 & 45 & $\infty$ & 3.63 & 2.54 & 3.52 & 0.08 & $\infty$ \\
G-14 & 3.0 & 1.4 & 0.63 & 0.89 & 1.0 & 45 & 300 & 3.63 & 2.13 & 3.52 & 0.08 & 217.64 \\
G-15 & 3.0 & 1.4 & 0.63 & 0.89 & 1.0 & 45 & 100 & 3.63 & 2.05 & 3.52 & 0.08 & 104.63 \\
G-16 & 3.0 & 1.4 & 0.63 & 0.89 & 1.0 & 45 & 30 & 3.63 & 1.09 & 3.53 & 0.08 & 46.89 \\
G-17 & 3.0 & 1.4 & 0.63 & 0.89 & 1.0 & 45 & 10 & 3.63 & 0.13 & 3.51 & 0.12 & 22.54 \\
G-18 & 3.0 & 1.4 & 0.63 & 0.89 & 1.0 & 45 & 1 & 3.63 & 0.00 & 3.15 & 0.48 & 4.86 \\
G-19 & 3.0 & 1.4 & 0.63 & 0.89 & 1.0 & 60 & $\infty$ & 3.63 & 6.15 & 3.46 & 0.10 & $\infty$ \\
G-20 & 3.0 & 1.4 & 0.63 & 0.89 & 1.0 & 60 & 300 & 3.63 & 4.46 & 3.47 & 0.10 & 217.64 \\
G-21 & 3.0 & 1.4 & 0.63 & 0.89 & 1.0 & 60 & 100 & 3.63 & 3.55 & 3.47 & 0.12 & 104.63 \\
G-22 & 3.0 & 1.4 & 0.63 & 0.89 & 1.0 & 60 & 30 & 3.63 & 3.36 & 3.48 & 0.11 & 46.89 \\
G-23 & 3.0 & 1.4 & 0.63 & 0.89 & 1.0 & 60 & 10 & 3.63 & 0.00 & 3.61 & 0.02 & 22.54 \\
G-24 & 3.0 & 1.4 & 0.63 & 0.89 & 1.0 & 60 & 1 & 3.63 & 0.00 & 3.06 & 0.57 & 4.86 \\
\hline
\multicolumn{13}{c}{Hit-and-run impact suite: $v_{\rm imp}=1.25v_{\rm esc}$} \\
\hline
H-1 & 0.90 & 1.00 & 0.21 & 0.63 & 1.25 & 35 & $\infty$ & 1.11 & 0.70 & 1.03 & 0.07 & $\infty$ \\
H-2 & 0.90 & 1.00 & 0.21 & 0.63 & 1.25 & 35 & 100 & 1.11 & 0.49 & 1.03 & 0.07 & 99.56 \\
H-3 & 0.90 & 1.00 & 0.21 & 0.63 & 1.25 & 35 & 30 & 1.11 & 0.21 & 1.03 & 0.07 & 44.62 \\
H-4 & 0.90 & 1.00 & 0.21 & 0.63 & 1.25 & 35 & 10 & 1.11 & 0.03 & 1.03 & 0.08 & 21.45 \\
H-5 & 3.0 & 1.4 & 0.62 & 0.88 & 1.25 & 35 & $\infty$ & 3.62 & 0.89 & 3.35 & 0.26 & $\infty$ \\
H-6 & 3.0 & 1.4 & 0.62 & 0.88 & 1.25 & 35 & 100 & 3.62 & 0.34 & 3.35 & 0.27 & 104.63 \\
H-7 & 3.0 & 1.4 & 0.62 & 0.88 & 1.25 & 35 & 30 & 3.62 & 0.24 & 3.35 & 0.27 & 46.89 \\
H-8 & 3.0 & 1.4 & 0.62 & 0.88 & 1.25 & 35 & 10 & 3.62 & 0.00 & 3.36 & 0.26 & 22.54 \\
\enddata
\tablecomments{
$M_{\rm tar}$, $M_{\rm imp}$, $M_{\rm tot}$, $M_{\rm planet}$, and $M_{\rm esc}$ are in $M_\oplus$; $R_{\rm tar}$ and $R_{\rm imp}$ are in $R_\oplus$; and $M_{\rm disk}$ is in lunar masses, $M_L$. $M_{\rm planet}$ is the mass of the postimpact primary planet, $M_{\rm disk}$ is the final disk-candidate mass identified by the orbital classification described in Section~\ref{sec:disk_identification}, and $M_{\rm esc}$ is the escaping mass.
The impact velocity $v_{\rm imp}$ is given in units of the mutual escape velocity, $v_{\rm esc}$; $\beta$ is the impact angle in degrees; and $P_{\rm orb}$ is the planetary orbital period in days.
$R_{\rm Hill}/R_{\rm planet}$ uses the postimpact equatorial radius from the converged classification iteration as $R_{\rm planet}$. Isolated-control simulations are denoted by $P_{\rm orb}=\infty$ and have $R_{\rm Hill}/R_{\rm planet}=\infty$.
IDs beginning with G denote the graze-and-merge impact suite with $v_{\rm imp}=1.0v_{\rm esc}$. IDs beginning with H denote the hit-and-run impact suite with $v_{\rm imp}=1.25v_{\rm esc}$ and a common impact angle of $\beta=35^\circ$.
The isolated cases provide fixed-velocity controls without stellar tidal or rotating-frame perturbations. All finite-period simulations listed in this table have \(\Omega_z>0\) and correspond to the prograde configuration.
}
\end{deluxetable*}

We used WoMa \citep{ruiz2021effect} to construct hydrostatic, differentiated planetary models and to initialize the target--impactor configurations. Each body consisted of a 30\,wt\% iron core and a 70\,wt\% forsterite mantle by mass. We considered two total system masses, $M_{\text{tot}} \approx 1.1\,M_\oplus$ and $3.6\,M_\oplus$, representative of terrestrial planets and super-Earths. Both the target and impactor were initialized with zero intrinsic spin in the corotating frame. The corresponding impactor-to-total mass ratios were $\gamma \approx 0.19$ and 0.17, respectively.
The simulations were resolved with approximately $5\times10^5$--$2\times10^6$ particles, depending on the impact suite and total system mass; this resolution is sufficient to study impact dynamics with MFM \citep{deng2019primordial,deng2019enhanced,yuan2023moon}. The target and impactor masses and radii, together with the measured postimpact disk, planet, and escaping masses and the ratio $R_{\rm Hill}/R_{\rm planet}$, are listed in Tables~\ref{tab:fiducial_suites}--\ref{tab:retrograde_impact}.

In the near-escape-velocity impact suite, the impact velocity was set equal to the mutual escape velocity of the target--impactor pair, $v_{\text{imp}} = v_{\text{esc}}$. The impact angle $(\beta)$ is defined as the angle between the relative velocity vector and the line joining the centers of the target and impactor at contact, with \(\beta=0^\circ\) corresponding to a head-on collision. We varied the impact angle over $\beta = 45^\circ$ and $60^\circ$, and the orbital period of the target planet over $P_{\text{orb}} = 1$, 10, 30, 100, and 300 days. We also performed an isolated-control simulation without stellar perturbations, corresponding to $\Omega = 0$. The center of mass of the target--impactor system was placed at the origin of the corotating frame, and the initial separation was set to $R_{\rm target}+R_{\rm impactor}$.

These low-speed impacts are graze-and-merge impacts \citep{stewart2012collisions}, which are likely least affected by stellar perturbations because they involve a spatial extent of only a few planetary radii. We also conducted a suite of hit-and-run simulations with $v_{\text{imp}} = 1.25v_{\text{esc}}$ for $P_{\text{orb}} = 10$, 30, and 100 days, adopting a common impact angle of $\beta = 35^\circ$, similar to the hit-and-run impact scenario for Moon formation \citep{reufer2012hit}.

High-speed impacts may preferentially occur for close-in planets because encounter velocities are tied to the local Keplerian velocity \citep{lissauer1993growth,jackson2018constraints}. To capture this dependence, we set \(v_{\infty}=fv_{\rm Kep}\), where $f$ parameterizes random velocity excitation. The impact velocity, $v_{\text{imp}}$, was then obtained by combining $v_{\infty}$ with the mutual escape velocity of the target--impactor pair, $v_{\text{esc}}$, through energy conservation \citep{lissauer1993growth}, as detailed in Appendix~\ref{app:velocity}. At fixed $f$, shorter orbital periods correspond to larger Keplerian velocities and therefore to larger values of $v_{\text{imp}}/v_{\text{esc}}$ \citep{mustill2018dynamical}. We performed selected high-velocity simulations with $f=0.1$, 0.2, 0.3, and 0.4 to compare different levels of orbital excitation.

To isolate the effect of prograde versus retrograde configurations, we performed an additional subset of simulations for the $1.1\,M_\oplus$ system. The impact geometry was kept fixed such that the angular momentum of the impact-generated disk remained along the positive $z$-direction, and only the sign of $\Omega_z$ was reversed. We refer to simulations with $\Omega_z>0$ as prograde and those with $\Omega_z<0$ as retrograde. All other initial conditions remained identical to those of the corresponding prograde simulations. The retrograde subset included the near-escape-velocity impacts with $\beta=45^\circ$ at $P_{\rm orb}=10$, 30, 100, and 300 days, the fixed-velocity hit-and-run impacts with $v_{\rm imp}=1.25v_{\rm esc}$ at $P_{\rm orb}=10$, 30, and 100 days, and three selected high-speed impacts listed in Table~\ref{tab:retrograde_impact}.

\begin{figure*}[ht!]
\plotone{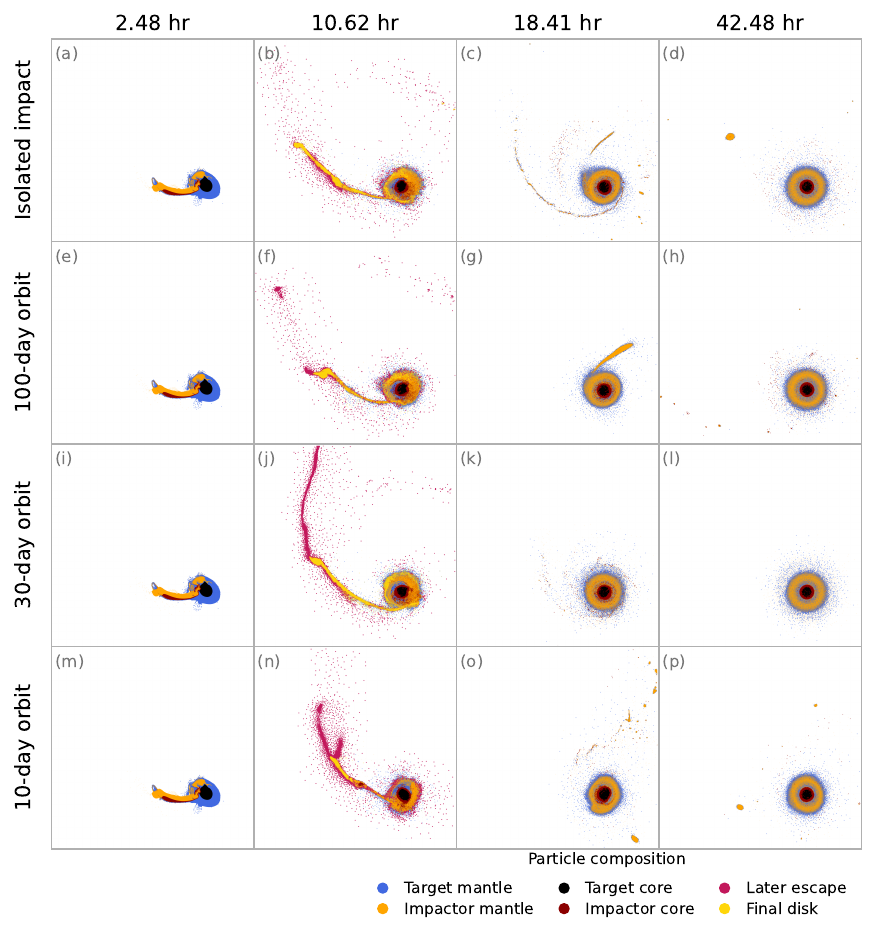}
\caption{Evolution of impact-generated debris under different levels of stellar perturbations for the fiducial impact setup with an impact angle of $45^\circ$, a total colliding mass of $1.1\,M_\oplus$, and an impact velocity of $v_{\rm imp}=v_{\rm esc}$. Rows show the isolated-control case and the $P_{\rm orb}=100$, 30, and 10~day cases; columns show snapshots at 2.48, 10.62, 18.41, and 42.48~hr after impact. Particles are shown in a midplane slice with $|z|<0.1\,R_\oplus$, and colors indicate material origin and composition. In the 10.62~hr snapshots, particles are additionally highlighted according to their final fate, identifying those incorporated into the disk and those that later escape from the system.}
\label{fig:morphology}
\end{figure*}

\subsection{Circumplanetary Disk Identification}
\label{sec:disk_identification}

We determined the postimpact disk properties using the iterative algorithm of \citet{canup2001scaling}. To account for tidal truncation of distant debris by the host star, we classified a particle as disk material only if it satisfied all of the following criteria: (1) it was gravitationally bound to the planet ($E<0$); (2) its periapsis lay outside the planet's equatorial radius; and (3) its apoapsis remained within the critical stability limit against stellar perturbations, $r_{\rm apo} < R_{\rm crit} \equiv \alpha R_{\rm Hill}$.

The Hill radius is defined as
\begin{equation}
R_{\rm Hill}
= a\left(\frac{M_{\rm planet}}{3M_\star}\right)^{1/3}.
\end{equation}
Here $a$ is the planet's orbital radius, $M_\star$ is the stellar mass (taken to be $1\,M_\odot$ throughout this work), and $M_{\rm planet}$ is the mass of the postimpact primary planet. Particles failing these criteria were classified as either escaping material or planet-bound material.

As established by \citet{holman1999long}, the critical boundary for orbital stability in high-mass-ratio systems scales linearly with the Hill radius. Following the empirical scaling of \citet{domingos2006stable}, we adopted $R_{\rm crit}=0.4895\,R_{\rm Hill}$ for the prograde simulations with $\Omega_z>0$ and $R_{\rm crit}=0.9309\,R_{\rm Hill}$ for the retrograde simulations with $\Omega_z<0$. For the isolated-control simulations, no stellar stability boundary was imposed.

Because the relaxation time of the postimpact debris varies with the initial impact conditions, the snapshot used for disk measurements was not fixed to a single simulation time. Instead, we selected the measurement snapshot once the large-scale radial oscillations of the debris had substantially decayed and the azimuthal velocity of most disk material had approached the local Keplerian velocity. All simulations were evolved for at least 35~hr after impact before disk properties were measured.

\begin{figure}[ht!]
\centering
\includegraphics[width=\linewidth]{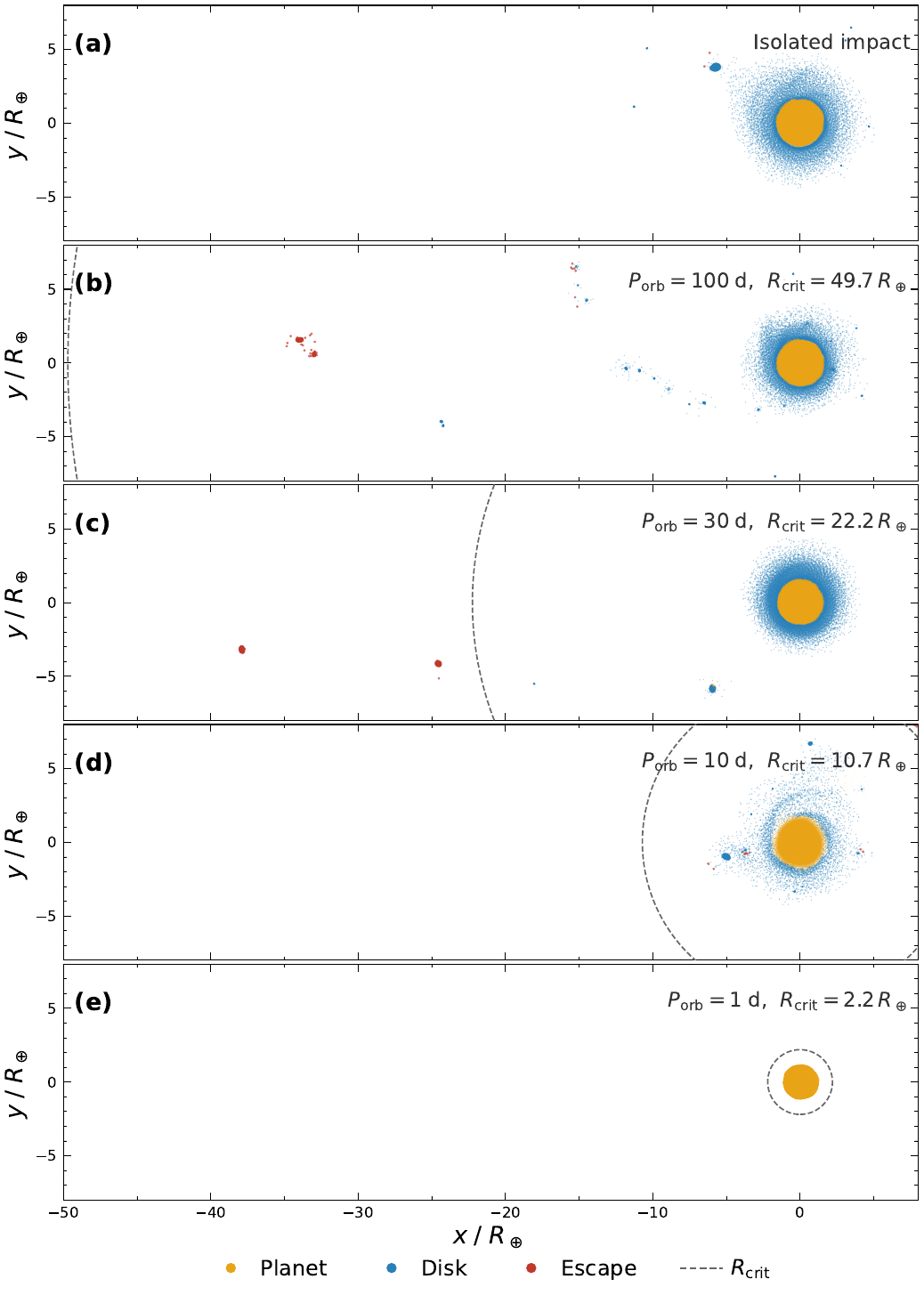}
\caption{Postimpact structure at 42.48~hr after impact for the fiducial setup with an impact angle of $45^\circ$, a total colliding mass of $1.1\,M_\oplus$, and $v_{\rm imp}=v_{\rm esc}$. Panels (a)--(e) show the outcomes for the isolated-control case and the $P_{\rm orb}=100$, 30, 10, and 1~day cases, respectively. Particles are shown in a midplane slice with $|z|<0.1\,R_\oplus$. Yellow points indicate planet-bound material, blue points indicate disk material, and red points indicate escaping material. The gray dashed curve marks the critical stability radius, $R_{\rm crit}=0.4895\,R_{\rm Hill}$.}
\label{fig:final}
\end{figure}

\section{Results}
\label{sec:results}
\subsection{Stellar Perturbations Drive Period-dependent Disk Depletion}

\begin{figure}[ht!]
\plotone{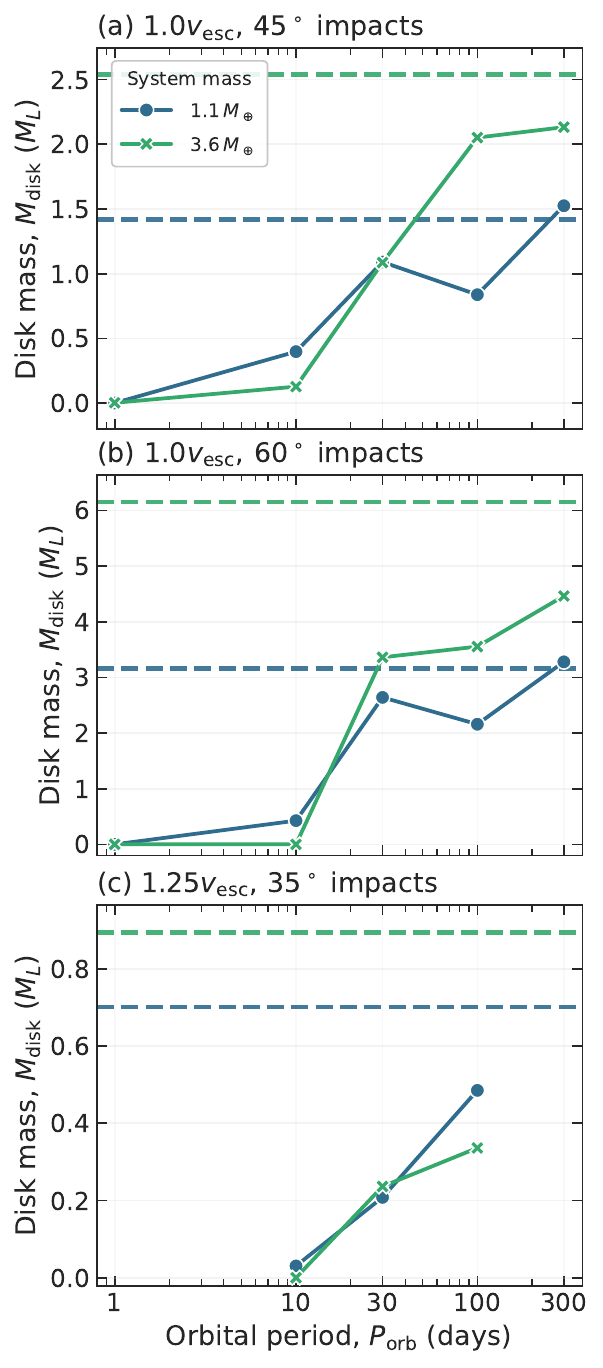}
\caption{Final disk mass as a function of orbital period. Panels (a) and (b) show impacts with angles of $45^\circ$ and $60^\circ$, respectively. Panel (c) shows hit-and-run impacts with $v_{\rm imp}=1.25v_{\rm esc}$ and an impact angle of $35^\circ$. The two curves correspond to systems with total colliding masses of $1.1\,M_\oplus$ and $3.6\,M_\oplus$. Dashed horizontal lines show the corresponding isolated-impact control values for comparison. Disk masses are expressed in lunar masses, $M_L$.}
\label{fig:disk_mass}
\end{figure}

Stellar tides can prevent impact ejecta from assembling into a CPD by removing high-angular-momentum debris from the planet's Hill sphere. In the isolated-control case, consistent with canonical Moon-forming giant-impact models \citep{canup2001origin,deng2019primordial}, disrupted impactor material remains largely bound, is sheared into a spiral arm, recollides with the target, and eventually settles into a CPD (Figure~\ref{fig:morphology}(a)--(d)). In Figure~\ref{fig:morphology}(d), an iron-depleted object reenters the visualized domain on an eccentric bound orbit, similar to the postimpact satellite reported in ultra-high-resolution SPH simulations \citep{kegerreis2022immediate}. However, the direct formation of postimpact satellites is sensitive to numerical resolution, algorithm, and impact configuration \citep{meier2025systematic}. For simplicity, we classify such clumps as part of the CPD when computing the total mass budget, and we ignore detailed clump--disk interactions (see Section~\ref{sec:disk_identification}).

\begin{figure*}[ht!]
\plotone{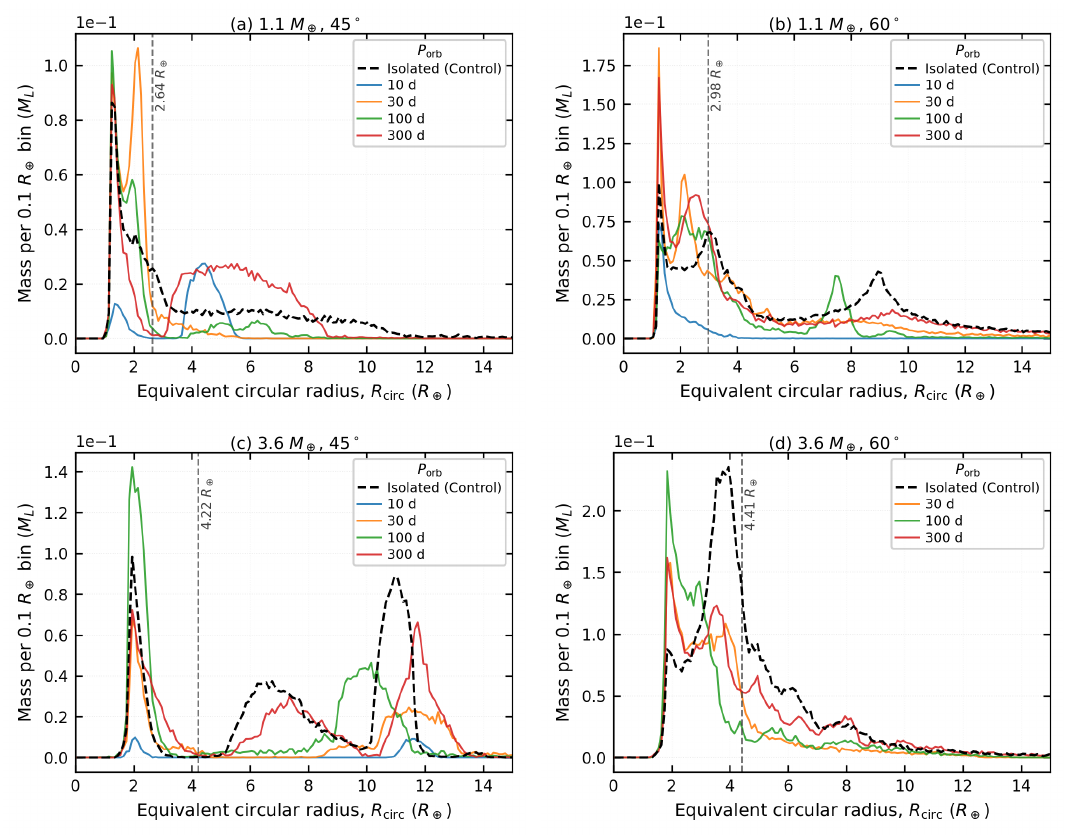}
\caption{Mass distribution of the surviving disk material as a function of equivalent circular radius $R_{\rm circ}$. The four panels show the $1.1\,M_\oplus$ and $3.6\,M_\oplus$ systems at impact angles of $45^\circ$ and $60^\circ$. For each particle, $R_{\rm circ}$ is computed from the prograde component of its specific angular momentum (see Equation~\ref{eq:rcirc}), and the debris mass is binned by $R_{\rm circ}$ (in lunar masses, $M_L$). Vertical dashed lines mark the Roche limits beyond which debris can agglomerate to form moons; these are calculated from the isolated-control model for each suite and shown as a common visual reference for each mass--impact-angle suite. Definitions of $R_{\rm circ}$ and the Roche-limit estimate are given in Appendix~\ref{app:rcirc_roche}; the resulting Roche-limit values and the corresponding Roche-interior and Roche-exterior disk masses and mass fractions are reported in Table~\ref{tab:roche_partition}.}
\label{fig:mass_distribution}
\end{figure*}

The early-stage evolution retains a similar spiral-arm morphology even as stellar perturbations increase (Figure~\ref{fig:morphology}). This is expected because the impact debris remains well within the Hill sphere, even for the $P_{\rm orb}=10$~day case (Table~\ref{tab:fiducial_suites}). In the isolated case, rotation of the spiral arm propels only a small fraction of diffuse material out of the system. However, when stellar perturbations are included, part of the spiral itself can be ejected as the loosely bound tail extends away from the postimpact planet (second column of Figure~\ref{fig:morphology}). At $P_{\rm orb}=30$~days, tidal forcing stretches and distorts the spiral arm significantly. At even shorter periods, stronger perturbations can modify the spiral-arm evolution and delay the merger of the impactor core into the target (Figure~\ref{fig:morphology}(n)).

The postimpact dynamical classification in Figure~\ref{fig:final} illustrates how stellar perturbations affect the mass and structure of the circumplanetary debris disk. In general, stronger perturbations around shorter-period planets suppress disk formation, yielding more compact and less massive disks. For planets on few-day orbits, no impact-generated CPD forms, so long-lived moons are not expected. This depletion is reflected in the final disk masses in Figure~\ref{fig:disk_mass}. Across both planetary masses and impact geometries, disk mass decreases toward shorter $P_{\rm orb}$: at $P_{\rm orb}=1$~day, all models produce zero disk mass. At $P_{\rm orb}=10$~days, disk formation remains inefficient, with final disk masses generally below $0.5\,M_L$. Even $\beta=60^\circ$ impacts, which can yield disks exceeding $3\,M_L$ at wider separations, produce only depleted disks at this period (Figure~\ref{fig:disk_mass}(b)).

Stellar perturbations may also place some debris onto intermediate orbits that remain within the Hill sphere (panels (b) and (c) of Figure~\ref{fig:final}), which can, in some cases, aid CPD formation. This likely contributes to the plateau in the disk-mass curves for some impact configurations (Figure~\ref{fig:disk_mass}).

The decrease in disk mass is typically associated with increasing escape mass for impacts onto closer-in planets (Table~\ref{tab:fiducial_suites}). However, it can also result from enhanced accretion. A striking example occurs for the $3.6\,M_\oplus$ planet at $P_{\rm orb}=10$~days and $\beta=60^\circ$, where the final disk mass is zero. Because the escaped mass is also nearly zero, the missing disk mass is not primarily lost from the system but is instead reaccreted by the planet during the simulated evolution. This case highlights a counterintuitive mass dependence of disk survival. At wide separations ($P_{\rm orb}\ge100$~days), the $3.6\,M_\oplus$ planets generally produce more massive disks than the $1.1\,M_\oplus$ planets, consistent with their larger mass reservoir. Near $P_{\rm orb}=10$~days, however, this ordering can reverse: the super-Earth models yield smaller disk masses than their lower-mass counterparts (Figure~\ref{fig:disk_mass}). This reversal marks a departure from the wide-orbit trend and motivates a closer examination of how stellar tides redistribute the surviving debris.

\subsection{Disk Structure}

We plot the mass distribution of the impact-formed CPDs as a function of equivalent circular radius, a proxy for specific angular momentum, in Figure~\ref{fig:mass_distribution}. The masses and corresponding fractions of disk material inside and outside the Roche limit are summarized in Table~\ref{tab:roche_partition}. As $P_{\rm orb}$ decreases, the disk mass distribution typically shifts to smaller equivalent circular radii, indicating preferential removal of high-angular-momentum material.
This trend echoes the increasingly compact CPDs around closer-in planets in Figure~\ref{fig:final}. These compact disks, especially those largely confined within the Roche radius, cannot form moons efficiently because only material outside the Roche radius can agglomerate \citep{ida1997lunar,salmon2012lunar}. The limited disk angular-momentum budget implies that only a small fraction of material can be transported beyond the Roche radius during subsequent disk evolution to support moon formation \citep{salmon2014accretion}.

Nevertheless, there are exceptions to this trend. In Figure~\ref{fig:mass_distribution}(c), the disk mass is dominated by several clumps, and the disk decreases in mass but does not become more compact as stellar perturbations increase. Clump formation and clump--disk interactions are sensitive to the numerical method and physical model \citep{meier2025systematic,salmon2012lunar}; a detailed investigation is beyond the scope of this paper. We therefore restrict this work to reporting the diverse postimpact properties rather than providing oversimplified estimates of exomoon masses \citep{ida1997lunar,salmon2014accretion}.

Alternatively, compact debris disks largely confined within the Roche limit could give rise to rocky planetary rings, although their subsequent evolution and long-term survival remain uncertain. Such rings could produce detectable signatures \citep{barnes2004transit,aizawa2018systematic,lane2026detecting}, potentially providing an additional observational tracer of close-in giant impacts.

\begin{figure*}[ht!]
\plotone{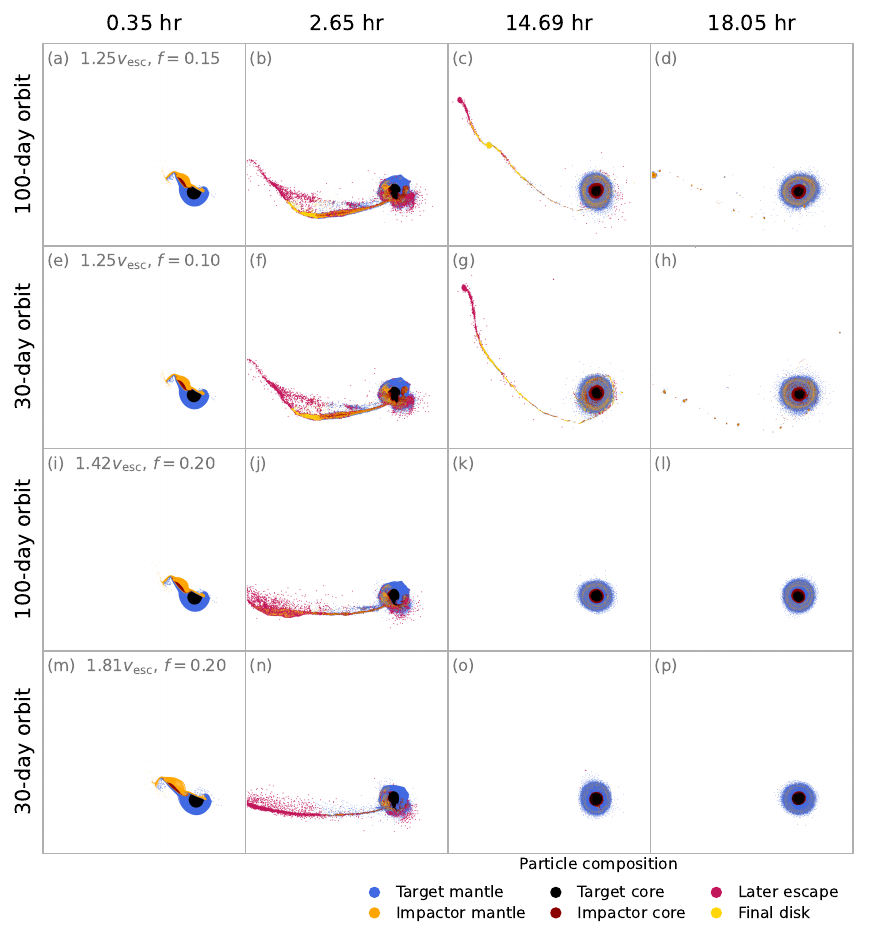}
\caption{Evolution of impact-generated debris for the high-velocity impact cases. Rows correspond to different impact scenarios with varying orbital periods ($P_{\rm orb}$) and impact velocities ($v_{\rm imp}$), as labeled. Columns show snapshots at 0.35, 2.65, 14.69, and 18.05~hr after impact. Particles are shown in a midplane slice with $|z|<0.1\,R_\oplus$, and colors indicate material origin and composition. In the 2.65 and 14.69~hr snapshots, particles are additionally highlighted according to their final fate, identifying those incorporated into the disk and those that later escaped from the system.}
\label{fig:highvel_morphology}
\end{figure*}

\begin{deluxetable*}{lccccccccccccc}
\tabletypesize{\scriptsize}
\tablewidth{0pt}
\tablecaption{Summary of the high-speed impact suite parameterized by the velocity-excitation factor $f$. \label{tab:high_f}}
\tablehead{
\colhead{ID} &
\colhead{$M_{\rm tar}$} &
\colhead{$R_{\rm tar}$} &
\colhead{$M_{\rm imp}$} &
\colhead{$R_{\rm imp}$} &
\colhead{$v_{\rm imp}$} &
\colhead{$f$} &
\colhead{$\beta$} &
\colhead{$P_{\rm orb}$} &
\colhead{$M_{\rm tot}$} &
\colhead{$M_{\rm disk}$} &
\colhead{$M_{\rm planet}$} &
\colhead{$M_{\rm esc}$} &
\colhead{$R_{\rm Hill}/R_{\rm planet}$}
}
\startdata
V-1 & 0.9 & 1 & 0.21 & 0.63 & 1.48 & 0.1 & 35 & 10 & 1.11 & 0.00 & 0.94 & 0.17 & 21.45 \\
V-2 & 0.9 & 1 & 0.21 & 0.63 & 1.42 & 0.2 & 35 & 100 & 1.11 & 0.05 & 0.95 & 0.16 & 99.56 \\
V-3 & 0.9 & 1 & 0.21 & 0.63 & 1.81 & 0.2 & 35 & 30 & 1.11 & 0.00 & 0.90 & 0.21 & 44.62 \\
V-4 & 0.9 & 1 & 0.21 & 0.63 & 2.39 & 0.2 & 35 & 10 & 1.11 & 0.00 & 0.86 & 0.25 & 21.45 \\
V-5 & 3.0 & 1.4 & 0.62 & 0.88 & 1.19 & 0.2 & 35 & 100 & 3.62 & 1.09 & 3.39 & 0.22 & 104.63 \\
V-6 & 3.0 & 1.4 & 0.62 & 0.88 & 1.39 & 0.2 & 35 & 30 & 3.62 & 0.00 & 3.18 & 0.44 & 46.89 \\
V-7 & 3.0 & 1.4 & 0.62 & 0.88 & 1.72 & 0.2 & 35 & 10 & 3.62 & 0.00 & 2.99 & 0.63 & 22.54 \\
V-8 & 3.0 & 1.4 & 0.62 & 0.88 & 1.40 & 0.3 & 35 & 100 & 3.62 & 0.00 & 3.18 & 0.44 & 104.63 \\
V-9 & 3.0 & 1.4 & 0.62 & 0.88 & 1.77 & 0.3 & 35 & 30 & 3.62 & 0.00 & 2.98 & 0.64 & 46.89 \\
V-10 & 3.0 & 1.4 & 0.62 & 0.88 & 2.32 & 0.3 & 35 & 10 & 3.62 & 0.00 & 2.86 & 0.76 & 22.54 \\
V-11 & 3.0 & 1.4 & 0.62 & 0.88 & 1.64 & 0.4 & 35 & 100 & 3.62 & 0.00 & 3.04 & 0.58 & 104.63 \\
V-12 & 3.0 & 1.4 & 0.62 & 0.88 & 2.18 & 0.4 & 35 & 30 & 3.62 & 0.00 & 2.89 & 0.73 & 46.89 \\
\enddata
\tablecomments{
Column definitions and units are the same as in Table~\ref{tab:fiducial_suites}, with the additional column $f$ denoting the velocity-excitation factor.
IDs beginning with V denote the orbit-coupled high-speed suite, in which the impact velocity depends on the local Keplerian velocity through the velocity-excitation factor $f$ (Appendix~\ref{app:velocity}). At fixed $f$, shorter orbital periods correspond to larger $v_{\rm Kep}$ and therefore larger $v_{\rm imp}/v_{\rm esc}$.
$R_{\rm Hill}/R_{\rm planet}$ uses the postimpact equatorial radius from the converged classification iteration as $R_{\rm planet}$. All high-speed simulations listed in this table have \(\Omega_z>0\) and correspond to the prograde configuration.
}
\end{deluxetable*}

\subsection{Velocity-enhanced Suppression}
\label{sec:highv}
\begin{figure}[ht!]
\plotone{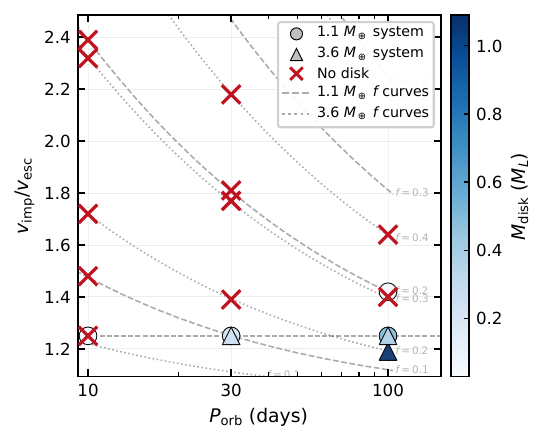}
\caption{Outcomes of the high-velocity impact suite as a function of orbital period and normalized impact velocity. Symbols represent individual simulations and are colored by the final disk mass $M_{\rm disk}$ (in lunar masses); red crosses indicate no surviving disk. The horizontal dashed line denotes the hit-and-run impact suite with $v_{\rm imp}=1.25v_{\rm esc}$, while the curves show orbit-coupled velocity tracks for different excitation factors $f$, defined by $v_\infty=fv_{\rm Kep}$. The definition of $f$ is given in Appendix~\ref{app:velocity}.}
\label{fig:high_vel}
\end{figure}

The preceding sections isolate the effect of stellar perturbations for near-escape-velocity impacts. We next examine whether this suppression strengthens as the impact velocity increases. This effect is particularly relevant for compact planetary systems because, for a fixed level of orbital excitation, the encounter velocity at infinity scales with the local Keplerian velocity. Shorter-period planets therefore reach larger values of $v_{\rm imp}/v_{\rm esc}$ for the same velocity-excitation factor $f$ (Appendix~\ref{app:velocity}). We use two comparisons to separate these effects: the hit-and-run suite in Table~\ref{tab:fiducial_suites}, which fixes $v_{\rm imp}=1.25v_{\rm esc}$ across orbital periods, and the orbit-coupled high-speed suite in Table~\ref{tab:high_f}, in which $v_\infty=fv_{\rm Kep}$.

The fixed-velocity hit-and-run suite provides a controlled extension of the near-escape-velocity impacts. In this suite, wider orbits still favor disk survival, whereas shorter-period cases retain only small or negligible disks (Figure~\ref{fig:disk_mass}(c)). At $P_{\rm orb}=10$~days, the terrestrial-mass system retains only a small disk, whereas the super-Earth system produces no surviving disk. Thus, even at a common impact velocity of $1.25v_{\rm esc}$, the period-dependent depletion seen in the near-escape-velocity suite persists. This comparison does not, however, include the additional velocity increase expected for close-in systems.

The orbit-coupled high-speed suite tests the same trend with impact velocities tied to the local orbital environment. The cases in Table~\ref{tab:high_f} sample $f=0.1$--$0.4$ and show that shorter orbital periods correspond to larger $v_{\rm imp}/v_{\rm esc}$ at fixed $f$. For example, in the terrestrial-mass cases with $f=0.2$, $v_{\rm imp}$ increases from $v_{\rm imp}=1.42v_{\rm esc}$ at $P_{\rm orb}=100$~days to $1.81v_{\rm esc}$ at 30~days and $2.39v_{\rm esc}$ at 10~days. The corresponding debris evolution in Figure~\ref{fig:highvel_morphology} shows that the high-velocity cases retain less coherent disk-forming material and produce a larger escaping component than the lower-velocity cases.

The full set of high-speed outcomes is summarized in Figure~\ref{fig:high_vel}. Nonzero disk masses are possible at relatively low impact velocities, whereas most simulations with $v_{\rm imp}\gtrsim1.4v_{\rm esc}$ produce little or no surviving disk. This behavior is consistent with previous giant-impact simulations \citep{leinhardt2012collisions,cambioni2019realistic}. Moderately faster hit-and-run impacts with $v_{\rm imp}\sim1.2$--$1.25v_{\rm esc}$ can still generate substantial disks \citep{reufer2012hit}, while the low-mass-ratio, high-velocity impacts surveyed by \citet{timpe2023systematic} produced disks smaller than $0.01\,M_L$ when $v_\infty>v_{\rm esc}$. More energetic hit-and-run collisions can instead produce substantial mantle stripping while allowing the smaller body to escape \citep{asphaug2014mercury}.

The $3.6\,M_\oplus$ case at $P_{\rm orb}=100$~days and $v_{\rm imp}=1.19v_{\rm esc}$ retains a disk of $1.09\,M_L$, which is more massive than the disk produced by the isolated impact with $v_{\rm imp}=1.25v_{\rm esc}$ (Table~\ref{tab:fiducial_suites}). However, super-Earth impacts in Table~\ref{tab:high_f} with $v_{\rm imp}>1.39v_{\rm esc}$ generate no disks, regardless of the strength of the stellar perturbations. The terrestrial-mass cases show the same tendency: the $P_{\rm orb}=100$~day model at $v_{\rm imp}=1.42v_{\rm esc}$ retains only $0.05\,M_L$, while the 30 and 10~day models at larger velocities retain no disk. Overall, high impact speeds strongly suppress disk-formation efficiency.

\begin{deluxetable}{lcccccccccccc}
\tabletypesize{\scriptsize}
\tablewidth{0pt}
\tablecaption{Retrograde impact subset.}
\label{tab:retrograde_impact}
\tablehead{
\colhead{ID} &
\colhead{$M_{\rm tar}$} &
\colhead{$R_{\rm tar}$} &
\colhead{$M_{\rm imp}$} &
\colhead{$R_{\rm imp}$} &
\colhead{$v_{\rm imp}$} &
\colhead{$\beta$} &
\colhead{$P_{\rm orb}$} &
\colhead{$M_{\rm tot}$} &
\colhead{$M_{\rm disk}$} &
\colhead{$M_{\rm planet}$} &
\colhead{$M_{\rm esc}$} &
\colhead{$R_{\rm Hill}/R_{\rm planet}$}
}
\startdata
R-1 & 0.90 & 1.00 & 0.21 & 0.65 & 1 & 45 & 300 & 1.11 & 1.29 & 1.08 & 0.02 & 207.09 \\
R-2 & 0.90 & 1.00 & 0.21 & 0.65 & 1 & 45 & 100 & 1.11 & 1.73 & 1.07 & 0.02 & 99.56 \\
R-3 & 0.90 & 1.00 & 0.21 & 0.65 & 1 & 45 & 30 & 1.11 & 1.31 & 1.07 & 0.02 & 44.62 \\
R-4 & 0.90 & 1.00 & 0.21 & 0.65 & 1 & 45 & 10 & 1.11 & 0.81 & 1.08 & 0.02 & 21.45 \\
R-5 & 0.90 & 1.00 & 0.21 & 0.63 & 1.25 & 35 & 100 & 1.11 & 0.52 & 1.03 & 0.07 & 99.56 \\
R-6 & 0.90 & 1.00 & 0.21 & 0.63 & 1.25 & 35 & 30 & 1.11 & 0.33 & 1.04 & 0.07 & 44.62 \\
R-7 & 0.90 & 1.00 & 0.21 & 0.63 & 1.25 & 35 & 10 & 1.11 & 0.17 & 1.04 & 0.07 & 21.45 \\
R-8 & 0.90 & 1.00 & 0.21 & 0.63 & 1.42 & 35 & 100 & 1.11 & 0.55 & 0.95 & 0.15 & 99.56 \\
R-9 & 0.90 & 1.00 & 0.21 & 0.63 & 1.48 & 35 & 10 & 1.11 & 0.00 & 0.94 & 0.17 & 21.45 \\
R-10 & 0.90 & 1.00 & 0.21 & 0.63 & 1.81 & 35 & 30 & 1.11 & 0.00 & 0.90 & 0.21 & 44.62 \\
\enddata
\tablecomments{
IDs beginning with R denote retrograde simulations with \(\Omega_z<0\). Each retrograde simulation used the same target--impactor configuration as its corresponding prograde simulation, with only the sign of \(\Omega_z\) reversed. $M_{\rm tar}$, $M_{\rm imp}$, $M_{\rm tot}$, $M_{\rm planet}$, and $M_{\rm esc}$ are in $M_\oplus$; $R_{\rm tar}$ and $R_{\rm imp}$ are in $R_\oplus$; and $M_{\rm disk}$ is in lunar masses, $M_L$. $M_{\rm planet}$ is the mass of the postimpact primary planet, $M_{\rm disk}$ is the final disk-candidate mass identified by the orbital classification described in Section~\ref{sec:disk_identification}, and $M_{\rm esc}$ is the escaping mass. The impact velocity $v_{\rm imp}$ is given in units of the mutual escape velocity, $v_{\rm esc}$; $\beta$ is the impact angle in degrees; and $P_{\rm orb}$ is the planetary orbital period in days.
}
\end{deluxetable}

\subsection{Retrograde Disks as a Favorable Limiting Case}

Stellar perturbations suppress impact-generated moons by reducing both the protolunar disk mass and the angular-momentum content of the surviving debris. This suppression is further strengthened when close-in impacts are coupled to higher encounter velocities. We therefore tested whether these trends persist for different disk orientations, as planet-formation studies suggest a broad (approximately random) distribution of planet spin orientations \citep{kokubo2010formation}.

Notably, retrograde circumplanetary orbits remain stable over a larger fraction of the Hill sphere than prograde orbits, and marginally escaping retrograde particles can be retained longer in the restricted three-body problem \citep{hamilton1991orbital,domingos2006stable}. To explore this favorable limiting case for disk formation, we performed a subset of retrograde simulations in which the impact configuration was held fixed and only the sign of $\Omega_z$ was reversed. The disk angular momentum was therefore antialigned with the planetary orbital angular momentum, as described in Section~\ref{sec:impact_setup}.

Graze-and-merge retrograde cases again show decreasing CPD mass for closer-in planets (Table~\ref{tab:retrograde_impact}). In general, they retain more disk material than the corresponding prograde cases at $P_{\rm orb}=100$, 30, and 10~days. The most notable case occurs at $P_{\rm orb}=100$~days, where the retrograde disk mass rises to $1.73\,M_L$, exceeding both the prograde 100~day case ($0.84\,M_L$) and the isolated near-escape-velocity control ($1.42\,M_L$). At $P_{\rm orb}=30$~days, the retrograde disk mass decreases to $1.31\,M_L$, comparable to that of the 300~day prograde case, and then decreases further at 10~days. At the widest tested orbit ($P_{\rm orb}=300$~days), the retrograde disk mass is slightly smaller than the prograde value, highlighting the sensitivity of the outcome to complex postimpact dynamics.

The high-velocity cases show the same qualitative pattern: the retrograde configuration can partially compensate for disk depletion, but it cannot prevent disk suppression at shorter orbital periods and higher impact velocities. This effect is most apparent in the 100~day case with $v_{\rm imp}=1.42v_{\rm esc}$, for which the retrograde configuration retains $0.55\,M_L$, compared with only $0.05\,M_L$ in the corresponding prograde case. In the shorter-period orbit-coupled cases, however, the 10~day case with $v_{\rm imp}=1.48v_{\rm esc}$ and the 30~day case with $v_{\rm imp}=1.81v_{\rm esc}$ produce no surviving disk in either configuration.

\begin{figure}[ht!]
\plotone{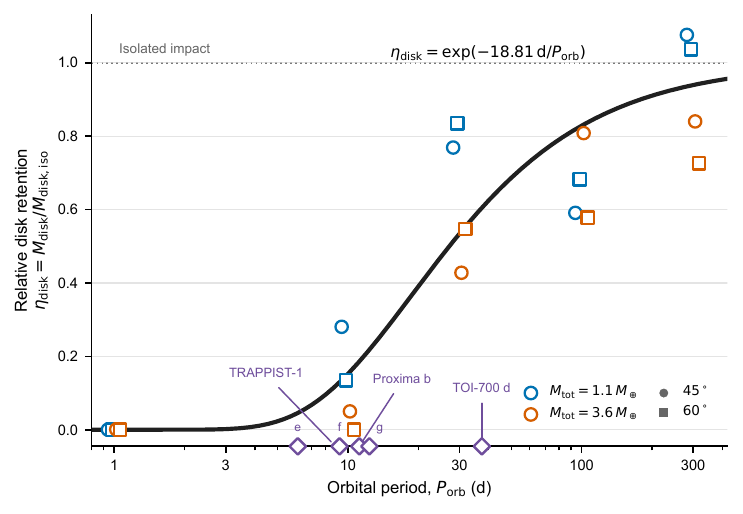}
\caption{Orbital-period dependence of the relative disk retention in the graze-and-merge suite. The retention fraction is defined as $\eta_{\rm disk}\equiv M_{\rm disk}/M_{\rm disk,iso}$. Blue and orange open symbols denote total colliding masses of $1.1\,M_\oplus$ and $3.6\,M_\oplus$, respectively, while circles and squares denote impact angles of $45^\circ$ and $60^\circ$. Simulation points at the same nominal orbital period are offset slightly in the horizontal direction for clarity. The solid black curve shows the empirical fit to the simulation results, and the gray dotted line marks the isolated-impact limit, $\eta_{\rm disk}=1$. Purple open diamonds along the horizontal axis mark the observed orbital periods of TOI-700 d \citep{gilbert2020first}, Proxima b \citep{anglada2016terrestrial}, and TRAPPIST-1 e--g \citep{gillon2017seven,agol2021refining}; the letters e, f, and g identify the individual TRAPPIST-1 planets.}
\label{fig:disk_retention_mdwarfs}
\end{figure}

\section{Discussion}
\label{sec:discussion}
\subsection{Orbital-period Scaling of Disk Retention and Implications for M-dwarf Systems}

M dwarfs provide a particularly relevant application because they constitute approximately 75\% of nearby stars \citep{winters2019solar} and frequently host close-in small planets \citep{dressing2015occurrence}. The low luminosities of M dwarfs place their habitable zones close to the host star, such that temperate terrestrial planets occupy short-period orbits \citep{kopparapu2013habitable}, where our simulations show substantial depletion of impact-generated disks.

The stellar perturbations included in our simulations are controlled by the orbital angular frequency rather than by the host-star mass itself (see Equation~\ref{eq:equation_of_motion} in Section~\ref{sec:method}). For a planet on a circular Keplerian orbit with \(M_{\rm planet}\ll M_\star\), the orbital angular frequency satisfies
\(\Omega^2=4\pi^2/P_{\rm orb}^2\simeq GM_\star/a^3\). The Hill radius can similarly be written as
\(R_{\rm Hill}\simeq(GM_{\rm planet}P_{\rm orb}^2/12\pi^2)^{1/3}\). Consequently, for otherwise identical planetary and impact conditions, our results apply to systems around stars of different masses. Nevertheless, we note that this local approximation is only valid when the impact occurs on a scale much smaller than $a$.

To quantify the disk mass retained relative to the corresponding isolated impact, we define the relative disk-retention factor as $\eta_{\rm disk}\equiv{M_{\rm disk}}/{M_{\rm disk,iso}}$, where \(M_{\rm disk,iso}\) is the disk mass produced by the corresponding isolated-impact control. For the graze-and-merge suite, its dependence on orbital period can be fitted by the empirical relation

\begin{equation}
\label{eq:disk_retention_fit}
    \eta_{\rm disk}
     =\exp\left(-\frac{18.81\,{\rm d}}{P_{\rm orb}}\right).
\end{equation}

To place our empirical scaling in the observational context \citep{pass2026jwst}, Figure~\ref{fig:disk_retention_mdwarfs} plots $\eta_{\rm disk}$ against the orbital periods of several representative temperate M-dwarf planets. The selected systems span the rapidly varying portion of the fitted relation. Among these systems, Proxima b and TRAPPIST-1 e--g occupy shorter-period orbits \citep{anglada2016terrestrial,gillon2017seven,agol2021refining}, for which our scaling implies lower disk retention under otherwise comparable impact conditions if such an impact occurs. TOI-700 d has the longest orbital period \citep{gilbert2020first} and therefore lies in a regime of comparatively weaker disk depletion. In addition, exomoons of habitable planets around M dwarfs may not be long-lived \citep{patel2026tidally}.

\subsection{Exomoon Formation in the Breaking-the-chains Scenario}

In planet-formation models, planetary embryos grow and migrate inward toward the inner edge of the gas disk, where convergent migration produces compact resonant chains \citep{izidoro2017breaking,dai2024prevalence}. As orbital damping weakens during gas-disk dispersal, resonant chains can become dynamically unstable, leading to giant impacts among the planets. This chain-breaking scenario may therefore lead to exomoon formation outcomes that differ from collisional growth models \citep{kokubo2010formation}.

Postdispersal $N$-body simulations produce collisions over a broad range of projectile-to-target mass ratios, including collisions between comparably massive planets \citep{esteves2022breaking}; such events are also consistent with the relatively uniform planetary masses inferred for compact resonant chains \citep{goldberg2022architectures}. Such collisions are of particular interest because larger impactor fractions are generally associated with more massive CPDs, although the angular-momentum budget and impact geometry remain important \citep{barr2017formation,timpe2023systematic,meier2025systematic,fang2025moon}.

The compact and nearly coplanar architecture of resonant chains may also bring planets into collision before prolonged scattering strongly excites their orbits \citep{esteves2020origins,goldberg2026close}. Such limited preimpact excitation may partly offset the amplification of impact velocities toward shorter orbital periods discussed in Section~\ref{sec:highv}. Together, comparable-mass colliders and limited preimpact excitation may favor disk formation. Chain breaking could therefore provide impact conditions conducive to producing massive exomoon-forming disks, motivating further investigation of the connection between planetary-system dynamics and giant impacts.

\section{Conclusion}
\label{sec:conclusion}
We performed multiple suites of giant-impact simulations of close-in rocky planets, incorporating stellar tidal and Coriolis forces. The outcomes are dramatically different from isolated moon-forming impacts: the stellar tidal environment can substantially suppress the generation of satellite-forming disks.

Close-in orbits reduce both the amount and the angular-momentum extent of bound circumplanetary material. The suppression is strongest for ultra-short-period and $\sim$10-day systems, where impact ejecta that would otherwise contribute to an extended disk are either removed from the stable circumplanetary region, reaccreted by the planet, or reorganized into compact structures rather than a mature satellite-forming disk. High-velocity impacts further reinforce this trend by shifting close-in collisions toward disk-poor outcomes. In contrast, retrograde impacts slightly ease this tension by producing more massive disks.

Thus, the apparent scarcity of close-in rocky exomoons may reflect not only observational limitations or subsequent tidal loss, but also an earlier bottleneck: the difficulty of assembling an extended satellite-forming disk in the first place. 

\begin{acknowledgments}
We are grateful to Bo Ma for stimulating discussions about the detection of exomoons. H.D. acknowledges support from the National Natural Science Foundation of China (NSFC) under grant No.~42550149. R.B. and Y.L. acknowledge support from the National Key Research and Development Program of China under grant No.~2024YFF0807500. C.R. acknowledges financial support from the Swiss National Science Foundation (SNSF, grants 200021\_207359 and 10001720).
\end{acknowledgments}

\restartappendixnumbering
\appendix

\section{Supplementary Methods}
\label{app:methods}


\subsection{Impact Velocity Parameterization}
\label{app:velocity}

Following the analytical framework of \citet{lissauer1993growth}, the encounter velocity at infinity can be parameterized as a fraction of the local Keplerian velocity,
\begin{equation}
v_{\infty} = f\,v_{\rm Kep},
\end{equation}
where $f$ represents the level of random velocity excitation associated with orbital eccentricities and inclinations, and $v_{\rm Kep}$ is the Keplerian velocity at the orbital radius of the target planet. The local Keplerian velocity is
\begin{equation}
v_{\rm Kep} = \sqrt{\frac{G M_\star}{a}},
\end{equation}
where $M_\star$ is the stellar mass and $a$ is the target planet's orbital radius, obtained from the orbital period through Kepler's third law.

The mutual escape velocity of the target--impactor pair is
\begin{equation}
v_{\rm esc} = \sqrt{\frac{2G\left(M_{\rm tar}+M_{\rm imp}\right)}{R_{\rm tar}+R_{\rm imp}}},
\end{equation}
where $M_{\rm tar}$ and $M_{\rm imp}$ are the target and impactor masses, and $R_{\rm tar}$ and $R_{\rm imp}$ are their corresponding radii.

The impact velocity at contact is then calculated from energy conservation as
\begin{equation}
v_{\rm imp} = \sqrt{v_{\rm esc}^{2} + v_{\infty}^{2}}.
\end{equation}
Equivalently, the impact velocity normalized by the mutual escape velocity is
\begin{equation}
\frac{v_{\rm imp}}{v_{\rm esc}} = \sqrt{1 + \left(f\,\frac{v_{\rm Kep}}{v_{\rm esc}}\right)^2}.
\end{equation}
At fixed $f$, shorter orbital periods correspond to larger $v_{\rm Kep}$ and therefore to larger encounter and impact velocities.

\subsection{Circularized Radius and Roche-limit Diagnostics}
\label{app:rcirc_roche}

To characterize the CPD material, we convert the disk-aligned component of the specific angular momentum into an equivalent circular radius. Because the positive $z$-direction is defined along the angular momentum of the impact-generated disk, we select particles with $j_z>0$ and define
\begin{equation}
\label{eq:rcirc}
R_{\rm circ} = \frac{j_z^2}{G M_{\rm planet}},
\end{equation}
where $M_{\rm planet}$ is the mass of the postimpact primary planet. We then bin the debris mass in $R_{\rm circ}$ and express the mass in each radial bin in units of lunar mass, $M_L$. This diagnostic maps the angular-momentum distribution of the surviving material onto a radial coordinate, allowing direct comparison with the planetary radius and the Roche limit under different tidal environments.

For the Roche limit shown in Figure~\ref{fig:mass_distribution}, we compute a thermodynamically informed value for each postimpact disk. This disk-specific estimate is adopted because the Roche boundary depends on the density of the material undergoing tidal disruption or accretion. In postimpact disks, that density is not well represented by a single prescribed silicate value, since the debris can contain a mixture of dense melt/solid material and an extended low-density vapor component, with the vapor fraction depending on the impact thermodynamics \citep[e.g.,][]{nakajima2014investigation,nakajima2022large}. Moreover, because satellite growth is ultimately fed by material that spreads across the Roche limit and becomes incorporated into condensed moonlets \citep{salmon2012lunar,salmon2014accretion}, we estimate the relevant disk density from the thermodynamic state of the surviving condensed debris.

Specifically, we define the fluid Roche limit using the classical expression \citep[e.g.,][]{murray1999solar}
\begin{equation}
a_R = 2.456\,R_{\rm planet}\left(\frac{\rho_{\rm planet}}{\rho_{\rm disk,cond}}\right)^{1/3},
\end{equation}
where $R_{\rm planet}$ and $\rho_{\rm planet}$ are the radius and mean density of the postimpact primary planet, respectively. The quantity $\rho_{\rm disk,cond}$ denotes the characteristic density of the condensed disk component. To estimate $\rho_{\rm disk,cond}$, we select disk particles with $\rho>2700\,\mathrm{kg\,m^{-3}}$ and compute their mass-weighted mean density. This threshold corresponds to the Tillotson-EOS reference density for basaltic/rocky material, which separates condensed and expanded EOS states in \citet{reinhardt2017numerical}. This procedure excludes the extended, low-density vapor component from the Roche-limit estimate. Table~\ref{tab:roche_partition} lists the resulting Roche limits, along with the disk masses and mass fractions inside and outside these limits for cases with $M_{\rm disk}\geq0.01\,M_L$.

\begin{deluxetable*}{lcccccccccccc}
\tabletypesize{\scriptsize}
\tablewidth{0pt}
\tablecaption{Roche-limit partition of the disk material. \label{tab:roche_partition}}
\tablehead{
\colhead{ID} &
\colhead{$M_{\rm tar}$} &
\colhead{$R_{\rm tar}$} &
\colhead{$M_{\rm imp}$} &
\colhead{$R_{\rm imp}$} &
\colhead{$v_{\rm imp}$} &
\colhead{$\beta$} &
\colhead{$P_{\rm orb}$} &
\colhead{$a_R/R_\oplus$} &
\colhead{$M_{\rm Roche,in}$} &
\colhead{$M_{\rm Roche,out}$} &
\colhead{$f_{\rm Roche,in}$} &
\colhead{$f_{\rm Roche,out}$}
}
\startdata
G-1  & 0.90 & 1.00 & 0.21 & 0.65 & 1.00 & 45 & $\infty$ & 2.64 & 0.64 & 0.78 & 0.45 & 0.55 \\
G-2  & 0.90 & 1.00 & 0.21 & 0.65 & 1.00 & 45 & 300 & 2.53 & 0.43 & 1.09 & 0.28 & 0.72 \\
G-3  & 0.90 & 1.00 & 0.21 & 0.65 & 1.00 & 45 & 100 & 2.77 & 0.69 & 0.14 & 0.83 & 0.17 \\
G-4  & 0.90 & 1.00 & 0.21 & 0.65 & 1.00 & 45 & 30 & 2.57 & 0.97 & 0.12 & 0.89 & 0.11 \\
G-5  & 0.90 & 1.00 & 0.21 & 0.65 & 1.00 & 45 & 10 & 2.91 & 0.07 & 0.32 & 0.19 & 0.81 \\
G-7  & 0.90 & 1.00 & 0.21 & 0.65 & 1.00 & 60 & $\infty$ & 2.98 & 0.96 & 2.20 & 0.30 & 0.70 \\
G-8  & 0.90 & 1.00 & 0.21 & 0.65 & 1.00 & 60 & 300 & 3.09 & 1.61 & 1.66 & 0.49 & 0.51 \\
G-9  & 0.90 & 1.00 & 0.21 & 0.65 & 1.00 & 60 & 100 & 2.82 & 1.09 & 1.07 & 0.50 & 0.50 \\
G-10 & 0.90 & 1.00 & 0.21 & 0.65 & 1.00 & 60 & 30 & 3.10 & 1.38 & 1.26 & 0.52 & 0.48 \\
G-11 & 0.90 & 1.00 & 0.21 & 0.65 & 1.00 & 60 & 10 & 2.79 & 0.38 & 0.04 & 0.90 & 0.10 \\
G-13 & 3.00 & 1.40 & 0.63 & 0.89 & 1.00 & 45 & $\infty$ & 4.22 & 0.51 & 2.03 & 0.20 & 0.80 \\
G-14 & 3.00 & 1.40 & 0.63 & 0.89 & 1.00 & 45 & 300 & 4.27 & 0.64 & 1.49 & 0.30 & 0.70 \\
G-15 & 3.00 & 1.40 & 0.63 & 0.89 & 1.00 & 45 & 100 & 4.39 & 1.00 & 1.05 & 0.49 & 0.51 \\
G-16 & 3.00 & 1.40 & 0.63 & 0.89 & 1.00 & 45 & 30 & 4.50 & 0.43 & 0.66 & 0.39 & 0.61 \\
G-17 & 3.00 & 1.40 & 0.63 & 0.89 & 1.00 & 45 & 10 & 4.62 & 0.04 & 0.08 & 0.34 & 0.66 \\
G-19 & 3.00 & 1.40 & 0.63 & 0.89 & 1.00 & 60 & $\infty$ & 4.41 & 3.70 & 2.45 & 0.60 & 0.40 \\
G-20 & 3.00 & 1.40 & 0.63 & 0.89 & 1.00 & 60 & 300 & 4.54 & 2.69 & 1.77 & 0.60 & 0.40 \\
G-21 & 3.00 & 1.40 & 0.63 & 0.89 & 1.00 & 60 & 100 & 4.57 & 2.81 & 0.75 & 0.79 & 0.21 \\
G-22 & 3.00 & 1.40 & 0.63 & 0.89 & 1.00 & 60 & 30 & 4.58 & 2.75 & 0.61 & 0.82 & 0.18 \\
\hline
H-1  & 0.90 & 1.00 & 0.21 & 0.63 & 1.25 & 35 & $\infty$ & 2.70 & 0.22 & 0.48 & 0.31 & 0.69 \\
H-2  & 0.90 & 1.00 & 0.21 & 0.63 & 1.25 & 35 & 100 & 2.63 & 0.12 & 0.36 & 0.26 & 0.74 \\
H-3  & 0.90 & 1.00 & 0.21 & 0.63 & 1.25 & 35 & 30 & 2.62 & 0.04 & 0.17 & 0.18 & 0.82 \\
H-4  & 0.90 & 1.00 & 0.21 & 0.63 & 1.25 & 35 & 10 & 2.97 & 0.01 & 0.02 & 0.43 & 0.57 \\
H-5  & 3.00 & 1.40 & 0.62 & 0.88 & 1.25 & 35 & $\infty$ & 4.04 & 0.58 & 0.32 & 0.65 & 0.35 \\
H-6  & 3.00 & 1.40 & 0.62 & 0.88 & 1.25 & 35 & 100 & 4.00 & 0.20 & 0.14 & 0.59 & 0.41 \\
H-7  & 3.00 & 1.40 & 0.62 & 0.88 & 1.25 & 35 & 30 & 4.10 & 0.14 & 0.09 & 0.61 & 0.39 \\
\hline
V-2  & 0.90 & 1.00 & 0.21 & 0.63 & 1.42 & 35 & 100 & 2.87 & 0.01 & 0.03 & 0.17 & 0.83 \\
V-5  & 3.00 & 1.40 & 0.62 & 0.88 & 1.19 & 35 & 100 & 3.82 & 0.33 & 0.76 & 0.30 & 0.70 \\
\hline
R-1  & 0.90 & 1.00 & 0.21 & 0.65 & 1.00 & 45 & 300 & 2.59 & 0.37 & 0.92 & 0.28 & 0.72 \\
R-2  & 0.90 & 1.00 & 0.21 & 0.65 & 1.00 & 45 & 100 & 2.60 & 0.75 & 0.98 & 0.44 & 0.56 \\
R-3  & 0.90 & 1.00 & 0.21 & 0.65 & 1.00 & 45 & 30 & 2.74 & 0.85 & 0.45 & 0.65 & 0.35 \\
R-4  & 0.90 & 1.00 & 0.21 & 0.65 & 1.00 & 45 & 10 & 3.16 & 0.12 & 0.69 & 0.15 & 0.85 \\
R-5  & 0.90 & 1.00 & 0.21 & 0.63 & 1.25 & 35 & 100 & 2.74 & 0.08 & 0.44 & 0.16 & 0.84 \\
R-6  & 0.90 & 1.00 & 0.21 & 0.63 & 1.25 & 35 & 30 & 2.64 & 0.05 & 0.29 & 0.15 & 0.85 \\
R-7  & 0.90 & 1.00 & 0.21 & 0.63 & 1.25 & 35 & 10 & 2.62 & 0.05 & 0.12 & 0.30 & 0.70 \\
R-8  & 0.90 & 1.00 & 0.21 & 0.63 & 1.42 & 35 & 100 & 2.91 & 0.02 & 0.25 & 0.07 & 0.93 \\
\enddata

\tablecomments{IDs beginning with G, H, V, and R denote the graze-and-merge, fixed-velocity hit-and-run, orbit-coupled high-speed, and retrograde suites, respectively. $M_{\rm tar}$ and $M_{\rm imp}$ are in $M_\oplus$; $R_{\rm tar}$ and $R_{\rm imp}$ are in $R_\oplus$; and $M_{\rm Roche,in}$ and $M_{\rm Roche,out}$ are in lunar masses, $M_L$. $v_{\rm imp}$ is in units of mutual escape velocity, $\beta$ is in degrees, and $P_{\rm orb}$ is in days. Only cases with $M_{\rm disk}\geq0.01\,M_L$ are listed.}
\end{deluxetable*}

\clearpage
\bibliography{reference}{}
\bibliographystyle{aasjournalv7.1}

\end{document}